\pdfoutput=1 %arXiv admin added this line
\documentclass[11pt,oneside,english]{amsart}
\usepackage[T1]{fontenc}
\usepackage[latin9]{inputenc}
\usepackage[letterpaper]{geometry}
\usepackage{textcomp}
\usepackage{amsthm}
\usepackage{graphicx}
\usepackage{setspace}
\usepackage{esint}
\makeatletter

\DeclareFontEncoding{LGR}{}{}
\DeclareTextSymbol{\~}{LGR}{126}

\numberwithin{equation}{section}
\numberwithin{figure}{section}
\theoremstyle{plain}
\newtheorem{thm}{Theorem}

\makeatother

\usepackage{babel}

\begin{document}

\title{\textmd{\textup{\normalsize Short-title: Theory on Time Required
for Vaccination }}}

\maketitle

\section*{\textbf{Theoretical Framework and Empirical Modeling for Time Required
to Vaccinate a Population in an Epidemic }}

\vspace{6cm}

\begin{center}

\author{\textbf{ARNI S.R. SRINIVASA RAO{*}}}

Indian Statistical Institute,

203 B.T. Road, Calcutta, INDIA 700108.

Email: arni@isical.ac.in.

Tel: +91-33-25753511.

\vspace{1cm}

\author{\textbf{KURIEN THOMAS}}

Department of Medicine

Christian Medical College, Vellore, India, 632004

Email: kurien123@hotmail.com

\vspace{4cm}

\end{center}

AMS subject classifications: 92D30, 60E05, 26D07

\vspace{0.4cm}

{*} Corresponding author

\vspace{0.2cm}

Acknowledgements: We are thankful to Lord Professor R.M. May, Oxford
for his encouragement. We thank Drs K Sudhakar (CDC, New Delhi), Damer
Blake (RVC, London), Professor N.V. Joshi (IISc, Bangalore) for their
valuable comments which improved our earlier drafts.

\pagebreak
\begin{abstract}
The paper describes a method to understand time required to vaccinate
against viruses in total as well as subpopulations. As a demonstration,
a model based estimate for time required to vaccinate H1N1 in India,
given its administrative difficulties is provided. We have proved
novel theorems for the time functions defined in the paper. Such results
are useful in planning for future epidemics. The number of days required
to vaccinate entire high risk population in three subpopulations (
villages, tehsils and towns) are noted to be 84, 89 and 88 respectively.
There exists state wise disparities in the health infrastructure and
capacities to deliver vaccines and hence national estimates need to
be re-evaluated based on individual performances in the states.
\end{abstract}


\section{Introduction}

In recent years vaccination against influenza A (H1N1) has become
one of the major concerns of health administrators around the globe.
H1N1 has now returned to prominence following its recurrence in India
and the news of 16 sudden deaths in the period between June and July,
2010 \cite{key-1}. Progress in Indian vaccine research has raised
the possibility of targeted or mass vaccination throughout India.
Importantly, the mere availability of a vaccine does not immediately
eliminate a pathogen from a population. Strategic planning covering
vaccine production, distribution and, if necessary, importation will
all be crucial components of a successful vaccination programme. India
is a vast country with a massively variable health infrastructure
distributed across Urban and Rural areas. Obtaining comprehensive
vaccination coverage in some parts of India will be a challenging,
but not impossible task. Recent precedents for the initiation of mass
anti-flu vaccinations include Canada, whose government has decided
to offer mass vaccination against H1N1 \cite{key-2} and, previously,
Israel who vaccinated its entire three million population against
flu in 1988 \cite{key-3}. Mathematical modeling can help to plan
vaccine strategies and designs in the event of H1N1 \cite{key-4}.
Modeling can also help in understanding the impact of population coverage
of H1N1 vaccines, impact in controlling due to delayed introduction
of vaccines \cite{key-5}. Mathematical analysis of the vaccination
and elimination of disease from population has been well studied \cite{new1,new2,new3,new4,new5}.
A completely new approach is introduced here in understanding the
time required to vaccinate whole population and sub-populations. The
concept of time required to vaccinate population and its sub-populations
is addressed in the paper is explained in sections 2 and 3. As an
example, we consider India and its states and divide these populations
by three sub-types (Village, Tehsil, Town) based on administrative
convenience. Although, a population can be divided into several other
types of subpopulations viz,. rural, urban or city, non-city urban,
town, capitals etc, it is divided into above three types because administratively,
such categorization helps in formal conceptualization of population
vaccination schemes in India. In addition, our abstract framework
proposed can be appropriately modified or expanded to suit several
countries administrative structure. Consideration of the breadth and
structure of the Indian population promotes the idea that vaccinating
only high risk subpopulations may be more effective in a resource
poor setting. Key questions associated with such large scale vaccination
programmes include: what will be the time required to vaccinate four
risk groups: pregnant women, children below five years of age, children
aged 6 to 15 years of age and health professionals? Similarly, how
long will be required to vaccinate the 1.20 billion population of
India? In response to these questions, can we project a realistic
timetable for effective vaccination using mathematical modeling to
inform government decision making (given a formal starting date)?
Answering these questions will help in assessing the potential epidemic
burden in the Indian population. The analysis described here will
be true for any emerging epidemic in India.

This paper is divided mainly into empirical modeling and theoretical
framework for understanding time required to vaccinate population
in the event of an epidemic. The resulting framework leads to new
results which provide bounds of \emph{time functions }introduced to
understand required time to cover the population after initiation
of population vaccination programme. These mathematical results can
be practically adoptable for visualizing strategies and their efficacy
in the health systems in the populations.

\section{Emperical model }

In this section we have computed time required to vaccinate high risk
population by dividing total population into various strata and then
carried out analysis through discrete computations.

Let $S_{i}(i=1,2,...,n_{1}),$ $T_{j}(j=1,2,...,n_{2})$ and $U_{k}(k=1,2,...,n_{3})$
be \emph{ith} tehsil, \emph{jth} town and \emph{kth} village in the
country. (Those who are not familier with these type of adminstrative
names, can think of three independent sub-populations. Entire population
is distributed into these three sub-populations). Let $V_{i}(t)$
be the number of the vaccinated population per $t$ units of time
in some location $l,$ $\alpha_{i}$ is the number of vaccine centers
per each type of location, $\beta_{i}$ is number of vaccinations
given per hour in a vaccine center in each type of location, $C_{i}$
is number of working hours for a vaccine center per day in each type
of location, $W_{i}$ is the number of working days per $t$ units
of time (Note : here $t$ units are per week or per month).

We calculate $V_{i}(t)$ by $V_{i}(t)=\alpha_{i}\beta_{i}C_{i}W_{i}(t)$.
Total population $P(t)$ is divided as $P(t)=p_{A}P(t)+p_{B}P(t),$
where $p_{A}$ and $p_{B}$ are the proportion of people in rural
and urban areas. Let $P_{S_{i}}(t),$ $P_{T_{j}}(t),$ $P_{U_{k}}(t)$
be populations of $ith$ tehsil, $jth$ town and $kth$ village respectively.
Then the total tehsil, town and village populations are

\begin{eqnarray}
P_{S}(t) & = & \sum_{i=0}^{n_{1}}P_{S_{i}}(t)\nonumber \\
P_{T}(t) & = & \sum_{j=0}^{n_{2}}P_{T_{j}}(t)\nonumber \\
P_{U}(t) & = & \sum_{k=0}^{n_{3}}P_{U_{k}}(t)\label{eq:1-1}\end{eqnarray}

We compute $V_{S}(t)$ , $V_{T}(t)$ , $V_{U}(t)$ and then required
time vaccination in the entire country by the type of location is
{\large $\frac{V_{S}(t)}{P_{S}(t)}$, $\frac{V_{T}(t)}{P_{T}(t)}$,
$\frac{V_{U}(t)}{P_{U}(t)}$}{\LARGE . }In case the populations are
very large, one can consider them as integral equations: $P_{S}(t)=\int_{0}^{n_{1}}P_{S_{i}}(t)di,$
$P_{T}(t)=\int_{0}^{n_{2}}P_{T_{j}}(t)dj,$ $P_{U}(t)=\int_{0}^{n_{3}}P_{U_{k}}(t)dk.$
Let $x_{ij}$, $y_{ik}$, $z_{il}$ be the times required to vaccinate
in the \emph{ith} state and \emph{jth} village, \emph{ith} state and
\emph{kth} tehsil, and \emph{ith} state and \emph{lth} town. We denote
$\underset{j}{max}(x_{ij}),$ $\underset{k}{max}(y_{ik}),$ $\underset{l}{max}(z_{il})$
for the corresponding maximum values of $ith$ rows and $\underset{j}{min}(x_{ij}),$
$\underset{k}{min}(y_{ik}),$ $\underset{l}{min}(z_{il})$ for the
corresponding mimimum values of $ith$ rows in the following matrices
$X$, $Y,$ $Z:$

$ $

$X=\left[\begin{array}{cccc}
x_{11}, & x_{12}, & ..., & x_{1V_{1}}\\
x_{21}, & x_{22}, & ..., & x_{2V_{2}}\\
\vdots & \vdots & \vdots & \vdots\\
x_{S1}, & x_{S2}, & ..., & x_{SV_{S}}\end{array}\right]$ , $Y=\left[\begin{array}{cccc}
y{}_{11}, & y_{12}, & ..., & y_{1H_{1}}\\
y_{21}, & y{}_{22}, & ..., & y_{2H_{2}}\\
\vdots & \vdots & \vdots & \vdots\\
y_{S1}, & y_{S2}, & ..., & y_{SH_{S}}\end{array}\right]$,

\begin{eqnarray}
\nonumber \\Z & = & \left[\begin{array}{cccc}
z_{11}, & z_{12}, & ..., & z_{1T_{1}}\\
z_{21}, & z_{22}, & ..., & z_{2T_{2}}\\
\vdots & \vdots & \vdots & \vdots\\
z_{S1}, & z_{S2}, & ..., & z_{ST_{S}}\end{array}\right]\label{eq:2}\end{eqnarray}

We assume that vaccination will be introduced simultaneously in all
three types of locations in each state. Then the time taken to completely
vaccinate the general population in each state $i$ is $\underset{}{max}\left\{ \underset{j}{max}(x_{ij}),\underset{k}{max}(y_{ik}),\underset{l}{max}(z_{il})\right\} $
for $j=1,2,...,V_{i};$ $k=1,2,...,H_{k};$ $l=1,2,...,T_{l}.$

Once we divide total centers in the country by the type of location
in which they exist, then the total high risk population living in
villages can be vaccinated within the time $\underset{i}{max}\left\{ \underset{i}{max}(x_{ij})\right\} $
for $i=1,2,...,S;$ $j=1,2,...,V_{i}.$ Similarly, the total risk
population living in tehsils and towns can be vaccinated within the
time $\underset{i}{max}\left\{ \underset{k}{max}(y_{ik})\right\} $
for $i=1,2,...,S;$ $k=1,2,...,H_{k}$ and $\underset{i}{max}\left\{ \underset{l}{max}(z_{il})\right\} $
for $i=1,2,...,S;$ $l=1,2,...,T_{i}.$

In case the vaccinations are introduced to the risk population in
the sequence of populations living in towns, tehsils, and villages
in the state $i$, then within the days of $\underset{i}{max}\left\{ \underset{l}{max}(z_{il})\right\} $
for $l=1,2,...,T_{i}$, $\underset{i}{max}\left\{ \underset{k}{max}(y_{ik})\right\} $
for $k=1,2,...,H_{i}$, $\underset{i}{max}\left\{ \underset{j}{max}(x_{ij})\right\} $
for $j=1,2,...,V_{i}$ the virus will spread to susceptible from infected
individuals. Hence, simultaneous introduction could reduce the overall
time required to vaccinate. This situation also impacts upon the development
of herd immunity. Mathematical analysis can help us to understand
the lower and upper limits of the time required to vaccinate in each
state. If we denote $R_{i}$ range of times for state $i$, then $R_{i}$
can be computed as $[L_{i},U_{i}],$ where

\begin{eqnarray}
L_{i} & = & \underset{i}{min}\left\{ \underset{j}{min}(x_{ij}),\underset{k}{min}(y_{ik}),\underset{l}{min}(z_{il})\right\} \nonumber \\
U_{i} & = & \underset{i}{max}\left\{ \underset{j}{max}(x_{ij}),\underset{k}{max}(y_{ik}),\underset{l}{max}(z_{il})\right\} \label{eq:3}\end{eqnarray}

The ranges of times taken for vaccinations in the towns and tehsils
can be computed by $R_{k}=[L_{k},U_{k}]$ and $R_{k}=[L_{k},U_{k}],$
where $L_{k}=\underset{i}{min}\left\{ \underset{k}{min}(y_{ik})\right\} $,
$U_{k}=\underset{i}{max}\left\{ \underset{k}{max}(y_{ik})\right\} $
and $L_{l}=\underset{i}{min}\left\{ \underset{l}{min}(z_{il})\right\} $,
$U_{l}=\underset{i}{max}\left\{ \underset{l}{max}(z_{il})\right\} .$

The above analytical description is not dependent on the number of
vaccine centers. We have provided arguments in this section for an
arbitrary size of vaccine centers allocated in towns, tehsils and
villages.

\section{Spatial Spread through convolution}

We introduce three functions $\{D_{i}(X),D_{j}(Y),D_{k}(Z)\}$, which
we call \emph{time functions }for three type of populations that we
are considering i.e. village, tehsil, and town in state $i.$ These
functions are defined as folllows:

\begin{eqnarray}
D_{i}(X) & = & \underset{j}{max}(x_{ij})-x_{ij'}\nonumber \\
D_{i}(Y) & = & \underset{k}{max}(y_{ik})-y_{ik'}\nonumber \\
D_{i}(Z) & = & \underset{l}{max}(z_{il})-z_{il'}\label{eq:4}\end{eqnarray}

where $x_{ij'}=\underset{i}{min}(x_{ij}),$ $y{}_{ik'}=\underset{i}{min}(y_{ik}),$
$z_{il'}=\underset{i}{min}(z_{il}).$ Observe that $\{min(x_{ij})\forall i\}$
is the set of minimum values of time taken to vaccinate villages in
all the states and that $\{max(x_{ij})\forall i\}$ is the set maximum
values of time taken to vaccinate villages in all the states. Let
$\mu$ be a measurable function describing the events $x_{ij'},y_{ik'},z_{il'}$
i.e. $\mu(x_{ij'})$, $\mu(y_{ik'})$, $\mu(z_{il'})$, are measurable
function for the minimum times, let $\sigma$ be the measurable function
describing the time functions $\{D_{i}(X),D_{j}(Y),D_{k}(Z)\}$ defined
as above. If we assume $\mu$ and $\sigma$ are Lebesgue integrable
on $(0,\infty)$, then convolution of $\mu$ and $\sigma$ is

\begin{eqnarray*}
\mu*\sigma & = & \int_{0}^{\infty}\mu(t-w)\sigma(w)dw\end{eqnarray*}
We also know that when $\mu$ and $\sigma$ are Lesbegue integrable
on the entire realline and at least one of $\mu$ or $\sigma$ is
bounded on the real line, then

\begin{eqnarray}
\mu*\sigma & = & \int_{-\infty}^{\infty}\mu(t-w)\sigma(w)dw\label{eq:5}\end{eqnarray}
 is bounded on the real line. Since the time taken to vaccinate is
bounded as the epidemic will not be lasting more than a season, the
property of convolution holds good. These kind of convolutions arise
in several applied mathematics areas apart from well known results
in pure mathematics (see \cite{key-14,key-15,key-17}). For recent
results applications of convolution approach see \cite{key-17} and
\cite{key-18}.

$\mu$ and $\sigma$ provides us an estimate of density function
of the maximum time taken to vaccinate in each state using convolution
approach. Since $\mu$ and $\sigma$ are Lebesgue integrable on $(0,\infty)$,
it is well-known that $\mu,\sigma\in L^{2}(0,\infty)$ {[}here $L^{2}(a,b)$
is the set of all real valued measurable functions $\mu$ , $\sigma$
on $(a,b)$ such that $\mu^{2}$, $\sigma^{2}$ are Lebesgue integrable
on $(a,b)].$ We can verify that for any real numbers $a_{1}$and
$a_{2}$, $a_{1}\mu+a_{2}\sigma$ is also in $L^{2}(0,\infty).$Hence
we can deduce $\left|(\mu,\sigma)\right|\leq\left\Vert \mu\right\Vert \left\Vert \sigma\right\Vert ,$
because inner product $(\mu,\sigma)$ is well defined by previous
statement. From these arguments, we can state following theorem for
the time functions.
\begin{thm}
Suppose $\mu(x_{ij}),$ $\mu(y_{k})$, $\mu(z_{il})$ are measurable
functions of the times taken to vaccinate and $\sigma(D_{i}(X)),$
$\sigma(D_{i}(Y))$, $\sigma(D_{i}(Z))$ are measurable functions
of the time functions as defined in the section, then

\begin{eqnarray*}
i)\left\Vert \mu(x_{ij'})+\sigma((D_{i}(X))\right\Vert  & \leq & \left\Vert \mu(x_{ij'})\right\Vert +\left\Vert \sigma((D_{i}(X))\right\Vert \end{eqnarray*}
 \\
\begin{eqnarray*}
ii)\left\Vert \mu(y_{ik'})+\sigma((D_{i}(Y))\right\Vert  & \leq & \left\Vert \mu(y_{ik'})\right\Vert +\left\Vert \sigma((D_{i}(Y))\right\Vert \end{eqnarray*}

\begin{eqnarray}
iii)\left\Vert \mu(z_{il'})+\sigma((D_{i}(Z))\right\Vert  & \leq & \left\Vert \mu(z_{il'})\right\Vert +\left\Vert \sigma((D_{i}(Z))\right\Vert \label{eq:6}\end{eqnarray}
\end{thm}
\begin{proof}
(i) follows by observing that,

$\left(\mu(x_{ij'}),\sigma((D_{i}(X))\right)$+$2\left(\mu(x_{ij'}),\sigma((D_{i}(X))\right)$+\begin{eqnarray*}
\left(\sigma((D_{i}(X)),\sigma((D_{i}(X))\right) & = & \left\Vert \mu(x_{ij'})\right\Vert ^{2}+\left\Vert \sigma((D_{i}(X))\right\Vert +2\left(\mu(x_{ij'}),\sigma((D_{i}(X))\right).\end{eqnarray*}
We can prove (ii) and (iii) by a similar arguement. \end{proof}
\begin{thm}
If \textup{$\underset{i}{min}(.)$ and $\underset{i}{max}(.)$ are
minimum and maximum values of the function '.' over set of all populations
$1\leq i\leq S$, and $x_{ij'}=\underset{i}{min}(x_{ij}),$ $x_{ij*}=\underset{i}{max}(x_{ij}),$
then }

\begin{eqnarray}
(i)\underset{i}{max}\left(D_{i}(X)\right) & \leq & \left|\underset{i}{min}(x_{ij'})-\underset{i}{max}(x_{ij*})\right|\nonumber \\
\nonumber \\(ii)\underset{i}{min}\left(D_{i}(X)\right) & \geq & \left|\underset{i}{min}(x_{ij*})-\underset{i}{max}(x_{ij'})\right|\label{eq:7}\end{eqnarray}

holds good.\end{thm}
\begin{proof}
\emph{(i) }Suppose $\underset{i}{max}\left(D_{i}(X)\right)$ attains
for the state $L$ for some $1\leq L\leq S.$ Let us denote this by
$D_{L}(X).$ Recall that we have information on set of all the times
taken for each sub-population within each state. Imagine for conceptual
clarity that we have plotted these times for each subpopulation vertically
on the $y-axis$ corresponding to the states in the $x-axis.$ Now,
let the coordinate corresponding to $L$ and at minimum of set of
time values is denoted by $P=(L,x_{Lj'})$ on the plane. The corresponding
co-ordinate on the set of maximum values obtained from each state
is denoted by $Q=(L,x_{Lj*})$ on the plane. Note that $x_{ij'}$
is the minumum value and $x_{ij*}$ is maximum value for the for state
$i.$

\emph{Case I. }Suppose $x_{Lj'}=\underset{i}{min}\left(x_{ij'}\right)=x_{i'j'}$
and $x_{Lj*}=\underset{i}{max}\left(x_{ij*}\right)=x_{i*j*}$, then

\begin{eqnarray*}
D_{L}(X) & = & \left|\underset{i}{min}(x_{ij'})-\underset{i}{max}(x_{ij*})\right|\end{eqnarray*}

\emph{Case II. }Suppose $x_{Lj'}\neq\underset{i}{min}\left(x_{ij'}\right)$
and $x_{Lj*}=\underset{i}{max}\left(x_{ij*}\right)=x_{i*j*}$, then
obviously $x_{Lj'}>\underset{i}{min}\left(x_{ij'}\right)=x_{i'j'}.$
Denote $P'=\left(i',x_{i'j'}\right).$ Let $\left(L,0\right)$ and
$\left(i',0\right)$be points on the $x-axis$ corresponding to the
states $L$ and $i'.$ Then, we have

\begin{onehalfspace}
\begin{eqnarray}
\left\Vert \left(L,0\right)-Q\right\Vert  & > & \left\Vert \left(L,0\right)-P\right\Vert >\left\Vert \left(i',0\right)-P'\right\Vert \label{eq:8}\end{eqnarray}

\begin{eqnarray*}
\implies\left|x_{Lj'}-x_{Lj*}\right| & < & \left|x_{i'j'}-x_{Lj*}\right|\end{eqnarray*}

\begin{eqnarray}
\implies D_{L}(X) & < & \left|\underset{i}{min}(x_{ij'})-\underset{i}{max}(x_{ij*})\right|.\label{eq:9}\end{eqnarray}

\end{onehalfspace}

\emph{Case III. }Suppose $x_{Lj'}=\underset{i}{min}\left(x_{ij'}\right)=x_{i'j'}$
and $x_{Lj*}\neq\underset{i}{max}\left(x_{ij*}\right)$. We have,
$x_{Lj*}<x_{i*j*}.$ Denote $Q*=\left(i*,x_{i'j*}\right).$ Since
$(i*,0)$ is a point on $x-axis$, it follows that $x_{i*j*}>x_{Lj*}$
and

\begin{spacing}{1.3}
\begin{eqnarray}
\left\Vert \left(i*,0\right)-Q'\right\Vert  & > & \left\Vert \left(L,0\right)-Q\right\Vert \label{eq:10}\end{eqnarray}

\begin{eqnarray*}
\implies\left|x_{i*j*}-x_{Lj'}\right| & > & \left|x_{Lj*}-x_{Lj'}\right|\end{eqnarray*}

\begin{eqnarray*}
\implies D_{L}(X) & < & \left|\underset{i}{min}(x_{ij'})-\underset{i}{max}(x_{ij*})\right|.\end{eqnarray*}

\end{spacing}

\emph{Case IV. }Suppose $x_{Lj'}\neq\underset{i}{min}\left(x_{ij'}\right)=x_{i'j'}$
and $x_{Lj*}\neq\underset{i}{max}\left(x_{ij*}\right)=x_{i*j*}$.
Although $Q*$ is the point corresponding to $x_{i'j*}$ and $P'=\left(i',x_{i'j'}\right)$
is the point corresponding to $x_{i'j'}$, we have,

\begin{singlespace}
\begin{eqnarray*}
\left\Vert P*-Q*\right\Vert  & < & D_{L}(X)\end{eqnarray*}
 and also,

\begin{eqnarray*}
\left\Vert P'-Q'\right\Vert  & < & D_{L}(X)\end{eqnarray*}

Again,

\begin{eqnarray}
\left\Vert \left(i*,0\right)-Q*\right\Vert  & > & \left\Vert \left(L,0\right)-Q\right\Vert \label{eq:11}\end{eqnarray}
 and

\begin{eqnarray*}
\left\Vert \left(i',0\right)-P'\right\Vert  & < & \left\Vert \left(L,0\right)-P\right\Vert \end{eqnarray*}
 From these arguement, we arrive at the following inequality,

\begin{eqnarray*}
D_{L}(X) & < & \left|x_{i'j'}-x_{i*j*}\right|\end{eqnarray*}

\end{singlespace}

i.e. $D_{L}(X)<\left|\underset{i}{min}(x_{ij'})-\underset{i}{max}(x_{ij*})\right|.$
Hence \emph{(i)} is proved.

\emph{(ii) }Suppose $\underset{i}{min}\left(D_{i}(X)\right)$ occurs
for state $\omega$ for some $1\leq\omega\leq S.$ Let $D_{\omega}(X)$
be the corresponding value. Corresponding to $D_{\omega}(X)$, let
us denote a point $U=\left(\omega,x_{\omega j'}\right)$ on the set
of values of minimum among $X$ in each state. The corresponding point
on the maximum values among $X$ in each state is $V=\left(\omega,x_{\omega j*}\right).$
Note that $x_{\omega j'}$ of $U$ and $x_{\omega j*}$ of $V$ need
not be minimum among set of all the minimum values and maximum among
set of all maximum values obtained for all the states. We will evaluate
the situation in four following cases.

\emph{Case I. }Suppose $x_{\omega j'}=\underset{i}{max}\left(x_{ij'}\right)=x_{i*j'}$
and $x_{\omega j*}=\underset{i}{min}\left(x_{ij*}\right)=x_{i'j*}$,
then it is clear from previous type of agrement, that,

\begin{eqnarray}
D_{\omega}(X) & = & \left|x_{i'j'}-x_{i*j'}\right|\nonumber \\
\nonumber \\ & = & \left|\underset{i}{min}(x_{ij*})-\underset{i}{max}(x_{ij'})\right|\label{eq:12}\end{eqnarray}
\emph{Case II. }Suppose $x_{\omega j'}\neq\underset{i}{max}\left(x_{ij'}\right)$
and $x_{\omega j*}=\underset{i}{min}\left(x_{ij*}\right)=x_{i'j*}.$
Let $U'=\left(i*,x_{i'j*}\right).$ Clearly, $x_{\omega j'}<x_{i*j*}.$
We have,

$\left\Vert U-V\right\Vert <\left\Vert \left(\omega,0\right)-V\right\Vert \mbox{ and }$$\left\Vert (\omega,0)-U\right\Vert <\left\Vert \left(i*,0\right)-U'\right\Vert .$
Hence,

\begin{eqnarray*}
\left|x_{\omega j*}-x_{\omega j'}\right| & > & \left|x_{\omega j*}-x_{i*j'}\right|\end{eqnarray*}

\begin{eqnarray}
\implies D_{\omega}(X) & > & \left|\underset{i}{min}(x_{ij*})-\underset{i}{max}(x_{ij'})\right|\label{eq:13}\end{eqnarray}

\emph{Case III. }Suppose $x_{\omega j'}=\underset{i}{max}\left(x_{ij'}\right)=x_{i*j'}$
and $x_{\omega j*}\neq\underset{i}{min}\left(x_{ij*}\right).$ The
situation arises to $x_{\omega j*}>x_{i'j*}.$ Let $V*=\left(i',x_{i'j*}\right).$
The minimum value in the set of all maximum times occurs at the point
$V*.$ Clearly,

$\left\Vert (i',0)-V*\right\Vert <\left\Vert \left(\omega,0\right)-V\right\Vert $
and $\left\Vert U-V\right\Vert <\left\Vert \left(\omega,0\right)-V\right\Vert $.
Hence,

\begin{eqnarray*}
\left|x_{\omega j*}-x_{\omega j'}\right| & > & \left|x_{i'j*}-x_{\omega j'}\right|\end{eqnarray*}

\begin{eqnarray*}
\implies D_{\omega}(X) & > & \left|\underset{i}{min}(x_{ij*})-\underset{i}{max}(x_{ij'})\right|\end{eqnarray*}

\emph{Case IV. }Suppose $x_{\omega j'}\neq\underset{i}{max}\left(x_{ij'}\right)=x_{i*j'}$
and $x_{\omega j*}\neq\underset{i}{min}\left(x_{ij*}\right).$ We
will have, $x_{\omega j'}<x_{i*j'}$ and $x_{\omega j*}>x_{ij*}.$
Let us assume $i*$ at point $V*$ is not equal to $i'$ at $U'.$
Observe that,

$\left\Vert (i*,0)-V*\right\Vert <\left\Vert \left(\omega,0\right)-V\right\Vert $
and $\left\Vert (i',0)-U'\right\Vert >\left\Vert \left(\omega,0\right)-U\right\Vert $.
Hence,

\begin{eqnarray*}
\left|x_{\omega j'}-x_{\omega j*}\right| & > & \left|x_{i'j*}-x_{\omega j'}\right|\end{eqnarray*}

\begin{eqnarray*}
\implies D_{\omega}(X) & > & \left|\underset{i}{min}(x_{ij*})-\underset{i}{max}(x_{ij'})\right|\end{eqnarray*}

Alternatively, if we assume $i*$ at point $V*$ is equal to $i'$
at $U'$, then the required results is straight forward.\end{proof}
\begin{thm}
If \textup{$\underset{i}{min}(.)$ and $\underset{i}{max}(.)$ are
minimum and maximum values of the function '.' over set of all populations
$1\leq i\leq S$, and $y_{ik'}=\underset{i}{min}(y_{ik}),$ $y_{ik*}=\underset{i}{max}(y_{ik}),$
then }

\begin{eqnarray}
(i)\underset{i}{max}\left(D_{i}(Y)\right) & \leq & \left|\underset{i}{min}(y_{ik'})-\underset{i}{max}(y_{ik*})\right|\nonumber \\
\nonumber \\(ii)\underset{i}{min}\left(D_{i}(Y)\right) & \geq & \left|\underset{i}{min}(y_{ik*})-\underset{i}{max}(y_{ik'})\right|\label{eq:14}\end{eqnarray}

holds good.\end{thm}
\begin{proof}
It follows from the similar logic given in proof of the Theorem 2.
We will consider the variables $D_{i}(Y)$, $D_{L}(Y)$ and $D_{\omega}(Y).$ \end{proof}
\begin{thm}
If \textup{$\underset{i}{min}(.)$ and $\underset{i}{max}(.)$ are
minimum and maximum values of the function '.' over set of all populations
$1\leq i\leq S$, and $z_{il'}=\underset{i}{min}(z_{il}),$ $z_{il*}=\underset{i}{max}(z_{il}),$
then }

\begin{eqnarray}
(i)\underset{i}{max}\left(D_{i}(Z)\right) & \leq & \left|\underset{i}{min}(z_{il'})-\underset{i}{max}(z_{il*})\right|\nonumber \\
\nonumber \\(ii)\underset{i}{min}\left(D_{i}(Z)\right) & \geq & \left|\underset{i}{min}(z_{il*})-\underset{i}{max}(z_{il'})\right|\label{eq:15}\end{eqnarray}

holds good.\end{thm}
\begin{proof}
It follows from the similar logic given in proof of the Theorem 2.
We will consider the variables $D_{i}(Z)$, $D_{L}(Z)$ and $D_{\omega}(Z).$
\end{proof}
We will now explore relationship between \emph{local time functions}
and \emph{global time functions} of vaccination. \emph{Local time
functions} we associate with subpopulations within a population and
\emph{global time functions} we associate with population itself.
Define $\widetilde{D}(i)=U_{i}-L_{i}$ for each $i.$ Note, $L_{i}$
must be equal to at least one of the values of $x_{ij}$, $y_{ik}$,
$z_{il}$ and $U_{i}$ must be exactly equal to at least one of the
values of $x_{ij}$, $y_{ik}$, $z_{il}$ for $j=1,2,...,V_{i}$,
$k=1,2,...,H_{i}$, $l=1,2,...,T_{i}.$ From this definition, we can
deduce following $S$ number of inequalities:

\begin{eqnarray}
\left\{ \left(x_{1j*}-x_{1j'}\right),\left(y_{1k*}-y_{1k'}\right),\left(z_{1l*}-z_{1l'}\right)\right\}  & \leq & \widetilde{D}(1)\nonumber \\
\left\{ \left(x_{2j*}-x_{2j'}\right),\left(y_{2k*}-y_{2k'}\right),\left(z_{2l*}-z_{2l'}\right)\right\}  & \leq & \widetilde{D}(2)\nonumber \\
\vdots &  & \vdots\nonumber \\
\left\{ \left(x_{Sj*}-x_{Sj'}\right),\left(y_{Sk*}-y_{Sk'}\right),\left(z_{Sl*}-z_{Sl'}\right)\right\}  & \leq & \widetilde{D}(S)\label{eq:16}\end{eqnarray}

\begin{thm}
Suppose $\left(x_{1j*}-x_{1j'}\right)\neq\left(y_{1k*}-y_{1k'}\right)\neq\left(z_{1l*}-z_{1l'}\right).$
Then,

\begin{eqnarray*}
i)\widetilde{D}(i)=\left(x_{1j*}-x_{1j'}\right) & \mbox{if, and only if,} & x_{ij*}=U_{i}\mbox{ and }x_{ij'}=L_{i}.\end{eqnarray*}

\begin{eqnarray*}
ii)\widetilde{D}(i)=\left(y_{1k*}-y_{1k'}\right) & \mbox{if, and only if,} & y_{ik*}=U_{i}\mbox{ and }y_{ik'}=L_{i}.\end{eqnarray*}

\begin{eqnarray*}
iii)\widetilde{D}(i)=\left(z_{1l*}-z_{1l'}\right) & \mbox{if, and only if,} & z_{il*}=U_{i}\mbox{ and }z_{il'}=L_{i}.\end{eqnarray*}
\end{thm}
\begin{proof}
\textit{(i) Suppose $\widetilde{D}(i)=\left(x_{1j*}-x_{1j'}\right).$
}\textit{\emph{This implies}}\textit{ $U_{i}-L_{i}=\left(x_{1j*}-x_{1j'}\right)$.
}\textit{\emph{Using the hypothesis and \ref{eq:16}, we have }}$\left(y_{1k*}-y_{1k'}\right)<\widetilde{D}(i)$
and $\left(z_{1l*}-z_{1l'}\right)<\widetilde{D}(i).$ This implies
$y_{ik*}<U_{i},$ $y_{ik'}>L_{i}$ and $z_{il*}<U_{i},$ $z_{il'}>L_{i}.$
Hence, $x_{ij*}=U_{i}$ and $x_{ij'}=L_{i}.$ Other part is straightforward.
Similarly, we can prove \emph{(ii) }and \emph{(iii)}.
\end{proof}

\section{Results}

We have demonstrated the idea of how to utilize the time function defined
in section 3 both empirically (by applying on Indian data) and found
bounds of these functions theoretically. The larger the time function
for a population indicates the larger the duration to cover population
in the same population. Once the vaccination is introduced in a sub-population
(say, $Z$) then the value of time function would attain at most the
absolute difference between $z_{il*}$ and $z_{il'}$.

The size of the Indian population is expected to be 1.2 billion by
the time of the 2011 census, distributed across 28 states and 7 union
territories. Decentralizing the administration of vaccine kit distribution
to 5767 tehsils, 7742 towns and 608786 villages by appointing nodal
officers to each of these three sectors would reduce vaccination time.
Population age, size and gender structure in each of these sectors
are not uniform and they vary across the 640 districts within India
\cite{key-6}. If a vaccine center is set-up in every village then
358 million people can be vaccinated per week (assuming seven shots
per hour in a twelve hour day, seven day working week). Importantly,
it is clear that only a proportion of villages will able to host a
vaccination center. In India, approximately 888 million people live
in villages, suggesting that it would take about 37 weeks (or 259
days) to vaccinate the entire rural population (assuming one in fifteen
villages hosts a vaccine center, see Figure 1). Since the urban population
is smaller than the rural population the number of vaccine centers
required in the tehsil and town sectors will be much lower than will
be required in villages. The installation of ten vaccine centers in
each town and tehsil would allow 80 million people living in these
areas to be vaccinated per week. At this rate it would take about
four weeks (or 28 days) to vaccinate the entire urban population.
Crucially, we estimate that 284.9 million people in villages and 100.1
million people in urban areas will fall into the combined risk group
of children, pregnant women and health professionals. There were studies
which found support to the model based idea of vaccinating high risk
populations against H1N1 \cite{key-5,key-6}. We estimate that it
will take about 11.9 weeks 12.69 weeks and 12.53 weeks to vaccinate
the combined risk group in villages, tehsils and towns (Figure 1).
Figures 2 and 3 illustrate the projected number of weeks required
to vaccinate the entire population in rural and urban areas given
a varying number of vaccination centers. We can also develop a mathematical
model equivalent to these discrete computations. Should the provision
and distribution of vaccine kits prove to be limiting then medium
range predictions indicate that rural areas could be vaccinated within
18 weeks and urban areas within 20 weeks from a given start date assuming
optimal vaccination center availability. The time taken to establish
vaccination centers in rural and urban areas will become the limiting
factor.

\section{Conclusions}

Both empirical and theoretical result presented in section 3 are novel
and gave new insights to handle population vaccine programs. Recent
precedents for large scale public health programmes in India include
the introduction of hepatitis B vaccination in ten major states in
2008, where 50 \% of the target population was vaccinated. Similarly,
vaccination against Japanese Encephalitis, introduced in 2006, targeted
27 million children in 11 states and yielded almost 16 million immunizations
\cite{key-9}. Under the universal immunization programme the government
of India has prioritized vaccination against six diseases including
tuberculosis. The highest coverage rate for BCG was reported in the
third DLHS to be 86 \% during 2007-2008 (District Level Health Survey).
Nonetheless, the third round of the National Family Health Survey
(NFHS-III) \cite{key-10} indicated that immunization rates vary widely
across the Indian states (Figure 4), influenced by variable vaccine
procurement capacity and distribution. Based on these state-specific
vaccine coverage rates the number of days likely to be required to
vaccinate village, tehsil and town populations were adjusted (Figure
5). Identification of the underlying causes of these disparities in
state-specific vaccination rates will require further research, taking
into state infrastructure, health facilities, etc. See appendix for
the impact of vaccination through the difference between two epidemic
densities obtained from pre-vaccination era and post-vaccination era.
Vaccine safety related issues also could lead to state level variations,
for example studies conducted in elsewhere indicate distrust over
H1N1 vaccines \cite{key-11,key-12,key-13}. In order for these theoretical
considerations to be achieved in practice it is essential that the
government strengthens i) the rural health infrastructure, either
empowering existing public health centers (PHCs) or setting-up new
flu vaccine centers, ii) procurement and distribution of the required
vaccines based upon size of the population, location etc. and iii)
methods of identifying and reaching the high risk population in a
timely manner.

\section*{Appendix}

We now use the framework of \cite{key-19} for pre-vaccinated and
post-vaccinated incidence densities. We assume that vaccination reduces
the time to elimination of a pathogen from a population.

Let $\gamma_{j}(j=1,2,...,t_{1})$ represent the peaks of infection
densities in the year $j.$ Suppose the number of years without introduction
of the vaccination into the population is $t_{1}$. Let $\gamma_{K}*(K=t_{1}+1,t_{2}+1,...)$
be the peaks of infection densities for the year K after introduction
of the vaccines into the population. Let $m(j)$ and $m(k)$ be the
means of incidence densities pre and post vaccinated populations,
$f_{j}$ and $f_{k}$ be the corresponding density functions. Peak
annual infection densities before vaccine introduction into the population
is assumed to be higher than the peak of the corresponding densities
after introduction of vaccination in the year $t_{1}+1$. Mean distance
between peaks of incidence in $j$ in $t_{1}+1$ is $d(\gamma_{j},\gamma_{t_{1}+1})$
(say), then the mean $\bar{d}(\gamma_{j},\gamma_{t_{1}+1})$ over
all possible pre-vaccinated years is over all possible pre-vaccinated
years is {\Large $\sum_{j}\frac{d(\gamma_{j},\gamma_{t_{1}+1})}{t_{1}+1}$}.
(See \cite{key-19}).\begin{eqnarray*}
\\\\\\\\\end{eqnarray*}

\includegraphics[scale=0.7]{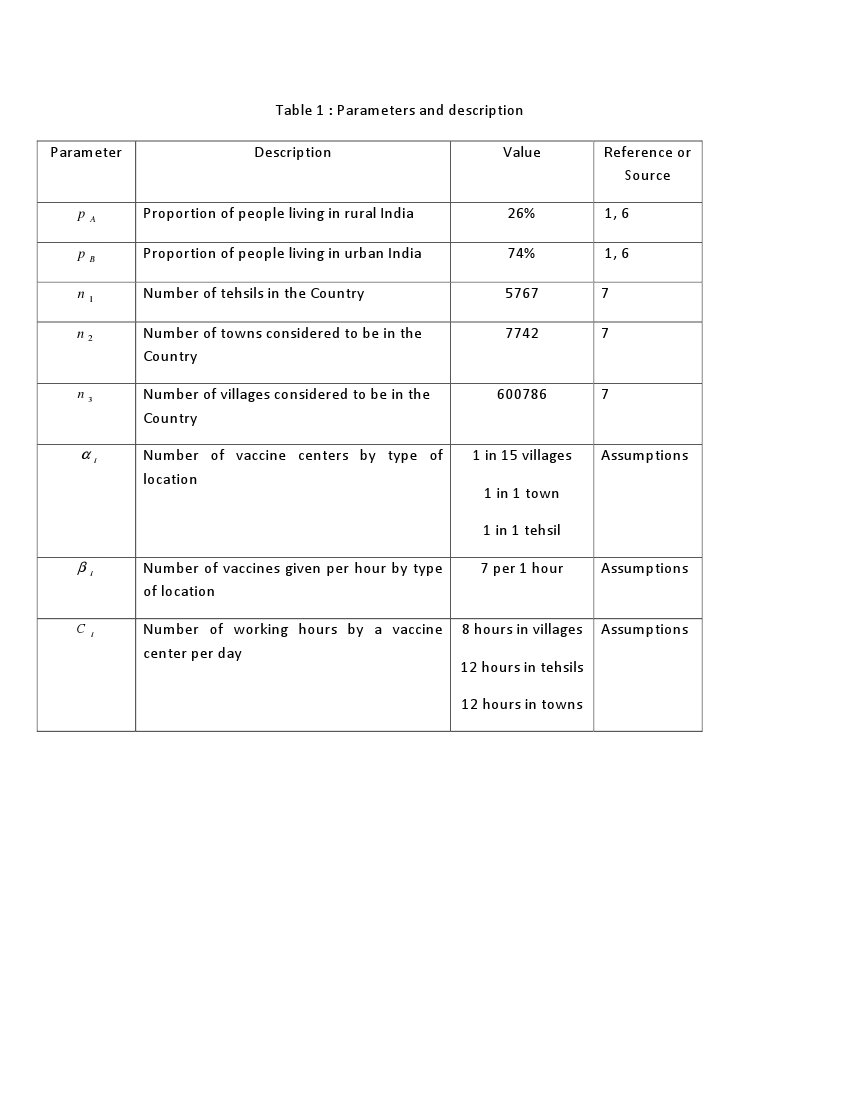}

\pagebreak

\includegraphics[scale=0.7]{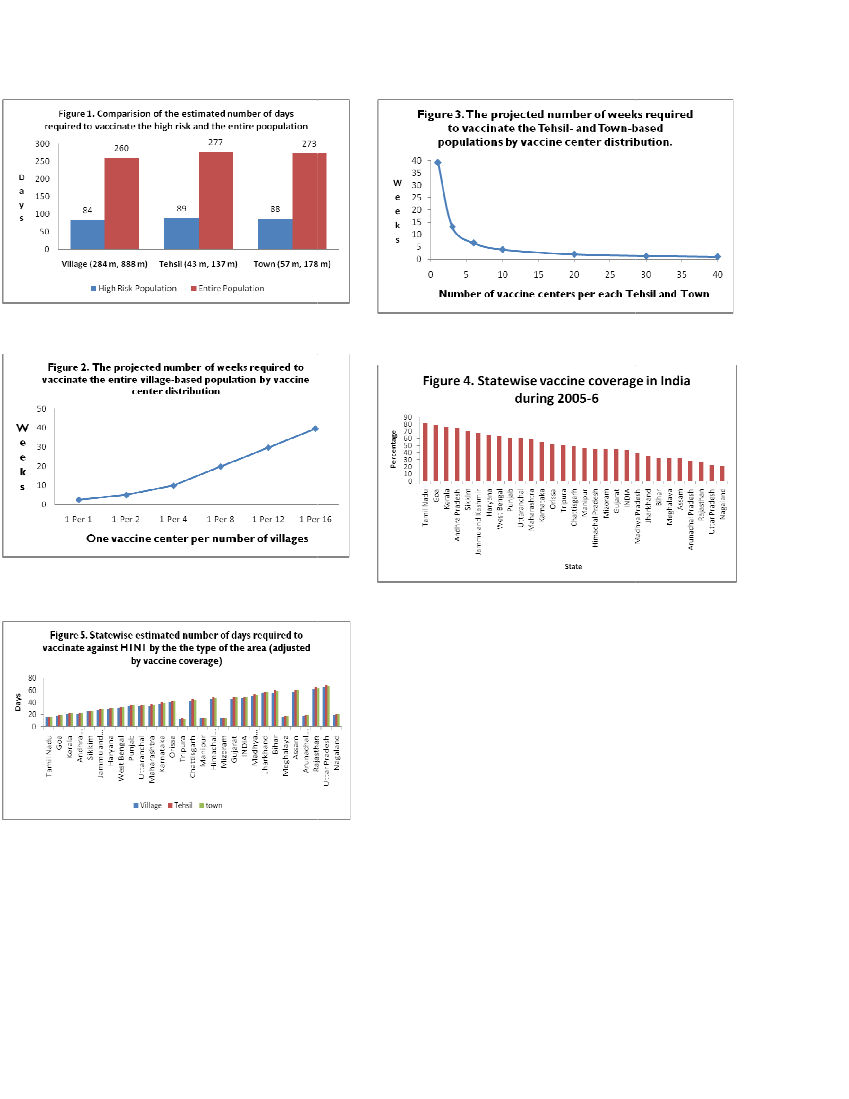}

\end{document}